\documentclass[a4paper,11pt]{article}
\usepackage{jcappub} 
\usepackage[normalem]{ulem}
\newcommand{\hinv}{{$h^{-1}$Mpc}}

\usepackage[style=numeric,backend=biber, sorting = none]{biblatex}

\arxivnumber{2311.14210} 
\title{\boldmath Self-calibrating BAO measurements in the presence of Small Displacement Interlopers}

\author[a,b]{Alan B. H. Nguyen}
\author[a,b]{Elena Massara}
\author[a,b,c]{Will J. Percival}

\affiliation[a]{Waterloo Centre for Astrophysics, University of Waterloo, 200 University Ave W, Waterloo, ON N2L 3G1, Canada}
\affiliation[b]{Department of Physics and Astronomy, University of Waterloo, 200 University Ave W, Waterloo, ON N2L 3G1, Canada}
\affiliation[c]{Perimeter Institute for Theoretical Physics,
31 Caroline St. North, Waterloo, ON N2L 2Y5, Canada}

\emailAdd{alan.nguyen@uwaterloo.ca}

\abstract{Baryon Acoustic Oscillation (BAO) observations offer a robust method for measuring cosmological expansion. However, the BAO signal in a sample of galaxies can be diluted and shifted by interlopers - galaxies that have been assigned the wrong redshifts. Because of the slitless spectroscopic method adopted by the Roman and Euclid space telescopes, the galaxy samples resulting from single line detections will have relatively high fractions of interloper galaxies. Interlopers with a small displacement between true and false redshift have the strongest effect on the measured clustering. In order to model the BAO signal, the fraction of such interlopers and their clustering need to be accurately known. We introduce a new method to self-calibrate these quantities by shifting the contaminated sample towards or away from us along the line of sight by the interloper offset, and measuring the cross-correlations between these shifted samples. The contributions from the different components are shifted in scale in this cross-correlation compared to the auto-correlation of the contaminated sample, enabling the decomposition and extraction of the component terms. We demonstrate the application of the method using numerical simulations and show that an unbiased BAO measurement can be extracted. Unlike previous attempts to model the effects of contaminants, self-calibration allows us to make fewer assumptions about the form of the contaminants such as their bias.}

\begin{document}
\maketitle
\flushbottom

\section{Introduction}
\label{sec:intro}

The upcoming Roman Space Telescope \cite{spergel2013widefield} and Euclid \cite{laureijs2011euclid} missions will soon provide astronomers with the next generation of space-based galaxy surveys. Both surveys will make use of slitless spectroscopy to obtain galaxy redshifts, where a grism in the optical path is used to disperse all incident light. While this method has the potential to quickly measure redshifts, it is not without its own sources of systematic error. The primary source of error we will study is the contamination of samples by interlopers arising due to line misidentification.

When measuring redshifts using emission lines, we compare observed and rest frame wavelengths, hence the measured redshift is dependant on the assumed rest frame wavelength. In most cases, this is not an issue since the use of multiple lines at known separations gives a secure result. However, to provide a larger sample for Roman and the baseline for Euclid we will include observations for which only a single line can be seen. This is where the sample can become contaminated. For example, [OIII] will be the primary line for high redshift Roman observations and is generally seen as a doublet \cite{wang2022}. However, when this doublet is unresolved into its two components, galaxies with strong H$\beta$ emission can be mistaken as [OIII] emitting galaxies with incorrect redshifts. Unfortunately for astronomers, the shift in position caused by the line misidentification is close to the BAO scale, and thus contaminates the BAO measurement \cite{massara2020}. It is crucial to correct for this systematic, otherwise the improved statistical error gained from the larger sample size is outweighed by the interloper contamination.

We focus on this type of contamination in Roman-like galaxy catalogues. \cite{foroozan2022} presented a method for modelling the auto-correlation of the contaminated sample that requires us to understand the [OIII]-H$\beta$ cross-correlation function. \cite{foroozan2022} assumed that this could be modelled directly using the small-scale [OIII] auto-correlation function, making the simplifying assumption for their analysis that the [OIII] and H$\beta$ biases were the same. Another method utilizing graph neural networks to estimate the fraction of interlopers was presented by \cite{massara2023}. We instead present a new method for estimating the galaxy target-interloper cross-correlation term based on measuring the cross-correlation between two replications of the contaminated catalogue, with the replications obtained by shifting the original catalogue along the line of sight (LOS) by $\pm 1 \times$ the interloper displacement. For simplicity and to demonstrate the method, we only consider the technique under the assumption of a global plane parallel LOS, approximating samples at large distance. The extension to wide, nearby, samples is left to future work. It is likely that this extension will degrade the sensitivity of our method as many of the features we see will lose their sharpness. This is discussed further in Section~\ref{sec:discussion}. Shifting the sample along the LOS, we change the scales of the target, interloper, and target-interloper cross-clustering terms compared to the corresponding terms in the auto-correlation of the contaminated sample. In this way, we may estimate the cross terms without any assumptions regarding the relative bias between targets and interlopers.

We can then use the measured cross-correlation between the two replicated catalogues in the model of the auto-correlation of the contaminated sample, based on fitting a linear model from CAMB \cite{lewis2000} coupled with nuisance parameters allowing for deviations from this model \cite{Ross-BOSS}. We use this model to find an unbiased estimation of the Alcock-Paczynski isotropic and anisotropic dilation parameters $\alpha$ and $\epsilon$ \cite{ap} from mock catalogues, in addition to the fraction of interlopers $f_{i}$.

Our paper is structured as follows. In Section~\ref{sec:interlopers}, we describe the effects of small displacement interlopers on the correlation function. In Section~\ref{sec:model}, we present our model for estimating the target-interloper cross term and the contaminated correlation function. In Section~\ref{sec:results}, we present results given by our pipeline. In Section~\ref{sec:discussion}, we discuss some limitations of the method and its application to real survey data. Finally, we conclude in Section~\ref{sec:conclusion}.

\section{The Effects of Small Displacement Interlopers}
\label{sec:interlopers}

\subsection{The Interloper Displacement}
\label{sec:displacement}
Interloper displacements are fixed in observed redshift, dependent on the offset in wavelength between the assumed and true lines. For a galaxy survey, we typically convert from redshifts to distances using a fiducial cosmological model, and so the interloper displacement in our map depends on the fiducial rather than the true cosmology. The comoving proper distance between an object with true redshift $z_{\text{true}}$ but displaced to $z_{\text{false}}$ is:
\begin{equation}
    \Delta_{\rm fid} = \int^{z_{\text{true}}}_{z_{\text{false}}} \frac{cdz}{H_{\rm fid}(z)} \approx \frac{c}{H_{\rm fid}(z_{\text{false}})}(z_{\text{true}} - z_{\text{false}})\,,
\end{equation}
where $H_{\rm fid}$ is the Hubble parameter in the fiducial cosmology, and $c$ is the speed of light. 
If we consider the case of [OIII]-H$\beta$ interlopers, this simplifies to
\begin{equation}
    \Delta_{\rm fid} \approx 87.41 \frac{1 + z_{\text{true}}}{(\Omega_{\Lambda,{\rm fid}} + \Omega_{m,{\rm fid}}(1 + z_{\text{true}})^{3})^{1/2}} \, h^{-1}\text{Mpc}\,,
\end{equation}
with $\Omega_{\Lambda,{\rm fid}}$ and $\Omega_{m,{\rm fid}}$ being the energy density of the cosmological constant and matter in the fiducial cosmology, respectively. \footnote{We remind the reader of the restframe wavelengths of [OIII] and H$\beta$: 500.7 nm and 486.1 nm.} For the range of redshifts of interest to Roman, z=1.0-2.8, and a Planck-like cosmology, this interloper shift ranges from $80\,h^{-1}$Mpc to $97\,h^{-1}$Mpc, quite close to the BAO feature. 

\subsection{The Contaminated Correlation Function}
\label{sec:contam_corrfunc}

With the introduction of interlopers into the target sample, the correlation function becomes ``contaminated'' as some fraction, $f_i$, of galaxies within the observed sample is displaced along the LOS. We will refer to the contaminated correlation function as $\xi_{cc}$. Furthermore, we will refer to the populations of interlopers and galaxy targets with the subscripts ``i" and ``g" respectively. The overdensity can then be written using two terms

\begin{equation}
    \delta_{c}(\Vec{x}) = (1 - f_{i}) \delta_{g}(\Vec{x}) + f_{i} \delta_{i}(\Vec{x})\,.\footnote{Here $\delta_c$ refers to the contaminated overdensity.} 
\end{equation}

Now we can model the contaminated correlation function as
\begin{align}
    \xi_{cc}(\Vec{r}, f_{i}) &= \langle \delta_{c}(\Vec{x_1})\delta_{c}(\Vec{x_2}) \rangle \\[0.7em] 
    &= \langle [(1 - f_{i}) \delta_{g}(\Vec{x_1}) + f_{i} \delta_{i}(\Vec{x_1})] [(1 - f_{i}) \delta_{g}(\Vec{x_2}) + f_{i} \delta_{i}(\Vec{x_2})] \rangle \\[0.7em] 
    &= (1 - f_i)^{2} \xi_{gg}(\Vec{r}) + f_{i}^{2} \xi_{ii}(\Vec{r}) + 2f_i(1-f_i) \xi_{gi}(\Vec{r})\,,
    \label{eq:cc_corr}
\end{align}

with $\Vec{r} = \Vec{x_2} - \Vec{x_1}$. Note the convention adopted follows that of \cite{foroozan2022}, and has the interloper catalogue already shifted along the LOS. I.e. the interlopers are considered at their observed rather than true positions.

Hereafter we will write the correlation functions as $\xi$ rather than $\xi(\Vec{r})$ for simplicity. The first two terms are the auto-correlations of galaxy targets and interlopers with an amplitude dependent on the interloper fraction. These terms contain most of the cosmological information and the familiar BAO peak. The third term is the cross-correlation between the galaxy targets and interlopers. When interlopers are at their true positions, there is a strong small scale clustering signal, as with galaxy-galaxy or interloper-interloper pairs. When interlopers are displaced along the LOS due to line confusion, this small scale clustering of the galaxy-interloper cross correlation is shifted to the displacement scale. This component typically skews the BAO peak in the contaminated catalogue clustering towards the interloper displacement and hence biases the standard BAO measurement.

\subsection{Adding Interlopers into Simulations}
\label{sec:interloper_add}

For the purposes of testing our model, we employ 1000 halo catalogues from the Quijote suite of N-body simulations \cite{quijotesims}. Our aim is to demonstrate the method, so the distribution of emission line galaxies in our simulations does not need to be realistic. We therefore choose not to populate the haloes with galaxies, but instead use the halo positions themselves. The simulations have a box length of 1~$h^{-1} {\rm Gpc}$ and use the fiducial Quijote suite parameters \cite{quijotesims}. In this configuration, the minimum halo mass is equal to $1.31\times10^{13}~h^{-1}$M$_\odot$. While it is true that we expect to find [OIII]-H$\beta$ interlopers in Roman-like catalogues mostly in redshift ranges greater than z = 1.8 \cite{wang2022}, we use the catalogue snapshots at redshift = 1. This range includes a greater density of objects, more consistent with the expected [OIII] mean density at z = 1.8 - 2.8 \cite{zhai2019}. We do use different interloper displacements corresponding to a redshift variation along the full Roman redshift range, as outlined in Section~\ref{sec:displacement}.

\begin{figure}[htbp]
\centering
\includegraphics[width=\textwidth]{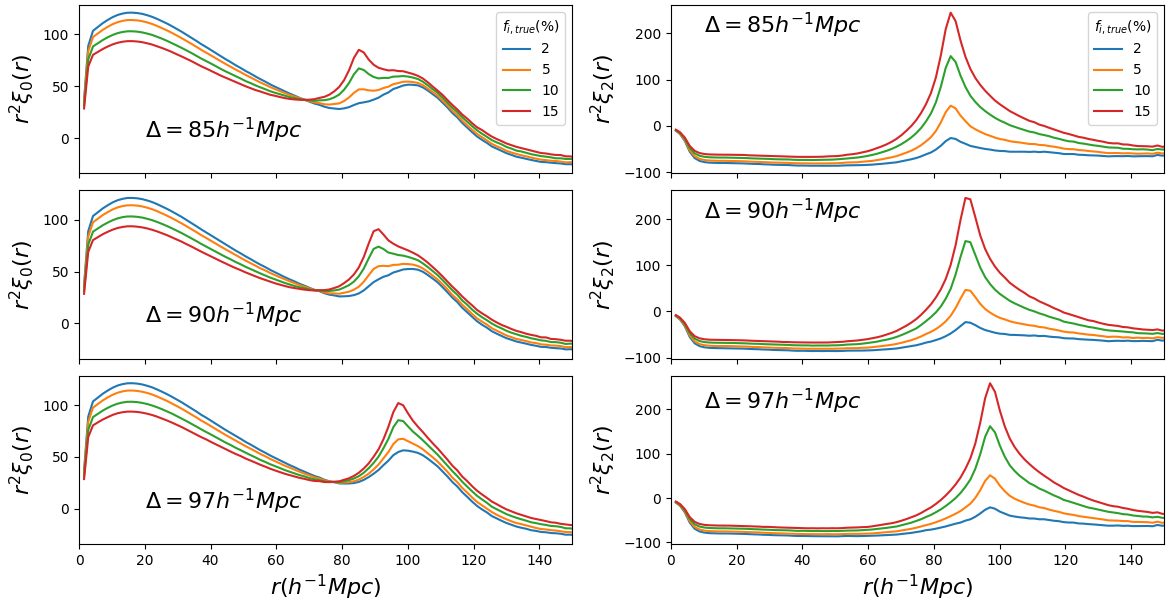}
\caption{The mean monopole (left) and quadrupole (right) of the correlation function measured from 1000 Quijote mocks for different interloper fractions and displacements. \label{fig:contaminated_corrfunc}}
\end{figure}

For each of the 1000 halo catalogues, we randomly-select a fraction $f_i$ of the total haloes, label them as interlopers, and shift them towards the observer along the LOS by a given shift $\Delta$. We generate catalogues with interloper fraction $f_i$ = (0.02, 0.05, 0.10, 0.15) and interloper shift $\Delta$ = (85, 90, 97) ${h}^{-1}{}\rm Mpc$. We then measure the correlation function using Nbodykit \cite{nbodykit} and 
the standard estimator,
\begin{equation}
   \label{eq:ddrr}
   \xi = \frac{DD}{RR} - 1,
\end{equation}
where $DD$ are the normalised catalogue-catalogue pairs and $RR$ are the expected random-random pairs calculated from the average halo density. We then 
extract the monopole, quadrupole, and hexadecapole moments. The effects of interlopers on the real-space monopole and quadrupole are shown in Figure~\ref{fig:contaminated_corrfunc}, showing the average of 1000 contaminated correlation function measurements with various interloper fractions and shifts. 

We also consider injecting interlopers into samples that also include redshift-space distortions. Here we move the full sample of haloes into redshift-space by shifting the line-of-sight displacements based on the halo velocities, before selecting a fraction of the haloes as interlopers. For simplicity we assume that all lines-of-sight are parallel {\bf ---} along one axis of the simulation. 

To illustrate the effect of differing galaxy biases between the galaxy target and interloper samples, we also generate a set of catalogues with a different selection of haloes for targets and interlopers. For these, all objects are ordered by mass and we select objects with the greatest mass as interlopers, down to the interloper fraction for the given catalogue. For example, for a catalogue with $f_i$ = 0.1, the top 10 percent most massive objects are selected as interlopers.

\section{Modeling the BAO Feature with Interlopers}
\label{sec:model}

\subsection{Overview of Method}
\label{sec:overview}

In order to model Equation~\ref{eq:cc_corr}, we need to estimate the galaxy target-interloper cross term accurately. Because of the interloper shift, the value of this term around the BAO scale comes from the small-scale target-interloper correlation function in true units (where interloper galaxies are at the true position), which are typically difficult to model due to nonlinearities.

We present an empirical method for measuring this term that uses the fact that we know $\Delta$ exactly and can use this to artificially apply further shifts to the contaminated catalogue, to create a new clustering measurement where the components that appear in Equation~\ref{eq:cc_corr} are further separated in scale.\footnote{The value of $\Delta$ is dependent on the assumption of an underlying fiducial cosmological model. The choice of fiducial model does not affect the method.} In this revised cross-correlation, the corrections for the nuisance terms can be modelled using the large-scale clustering, such that we do not require a theoretical model of the small-scale clustering signal. This concept works in redshift-space and for situations where target and interloper galaxies have different biases. 

The following subsections describe this process in detail. 

\begin{figure}[htbp]
\centering
\setlength{\fboxrule}{1pt}%
\fbox{\includegraphics[width=0.6\textwidth]
    {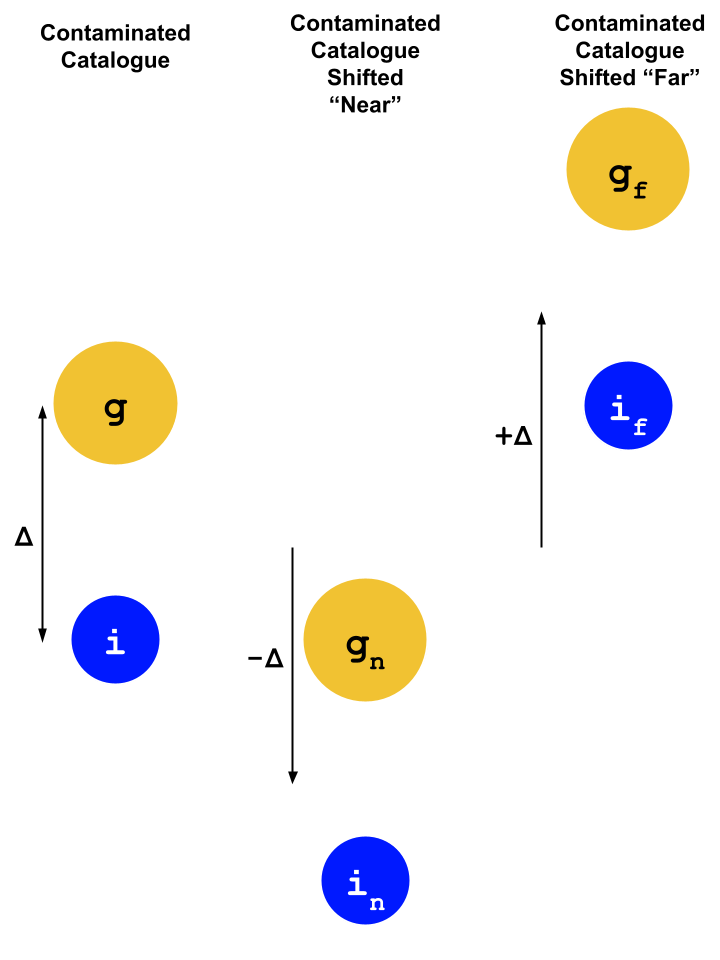}}
\caption{Schematic diagram of the contaminated galaxy catalogue on the left and the effect of shifting it to generate the near-shifted and far-shifted catalogues. Galaxy targets at their true positions are labelled with $g$ while interlopers are labelled with $i$. The interloper shift is $\Delta$. \label{fig:schematic}}
\end{figure}

\subsection{The Near-Far Shifted Correlation Function}
\label{sec:selfcalib}

We introduce two additional catalogues. The \textit{far-shifted} catalogue is the contaminated catalogue with all object positions shifted \textit{away} from the observer by the known interloper shift. The \textit{near-shifted} catalogue is the same contaminated catalogue with the object positions moved \textit{towards} the observer by one interloper shift. Each catalogue is still made up of galaxy targets and interlopers. A schematic diagram for this process is shown in Figure~\ref{fig:schematic}.

We use the cross correlation between the near-shifted and far-shifted catalogues:
\begin{equation}
    \xi_{c_{n}c_{f}} = (1 - f_i)^{2} \xi_{g_{n}g_{f}} + f_{i}^{2} \xi_{i_{n}i_{f}} + f_i(1-f_i) \xi_{g_{n}i_{f}} + f_i(1-f_i) \xi_{i_{n}g_{f}}\,,
    \label{eq:ud_corr}
\end{equation}
with the subscript $_{f}$ referring to \textit{far-shifted} components and the subscript $_{n}$ referring to \textit{near-shifted} components. Note that unlike the case in Equation~\ref{eq:cc_corr}, the target-interloper and interloper-target cross correlations are now not equal and thus cannot be combined into a single term. 

\begin{figure}[htbp]
\centering
\includegraphics[width=\textwidth]{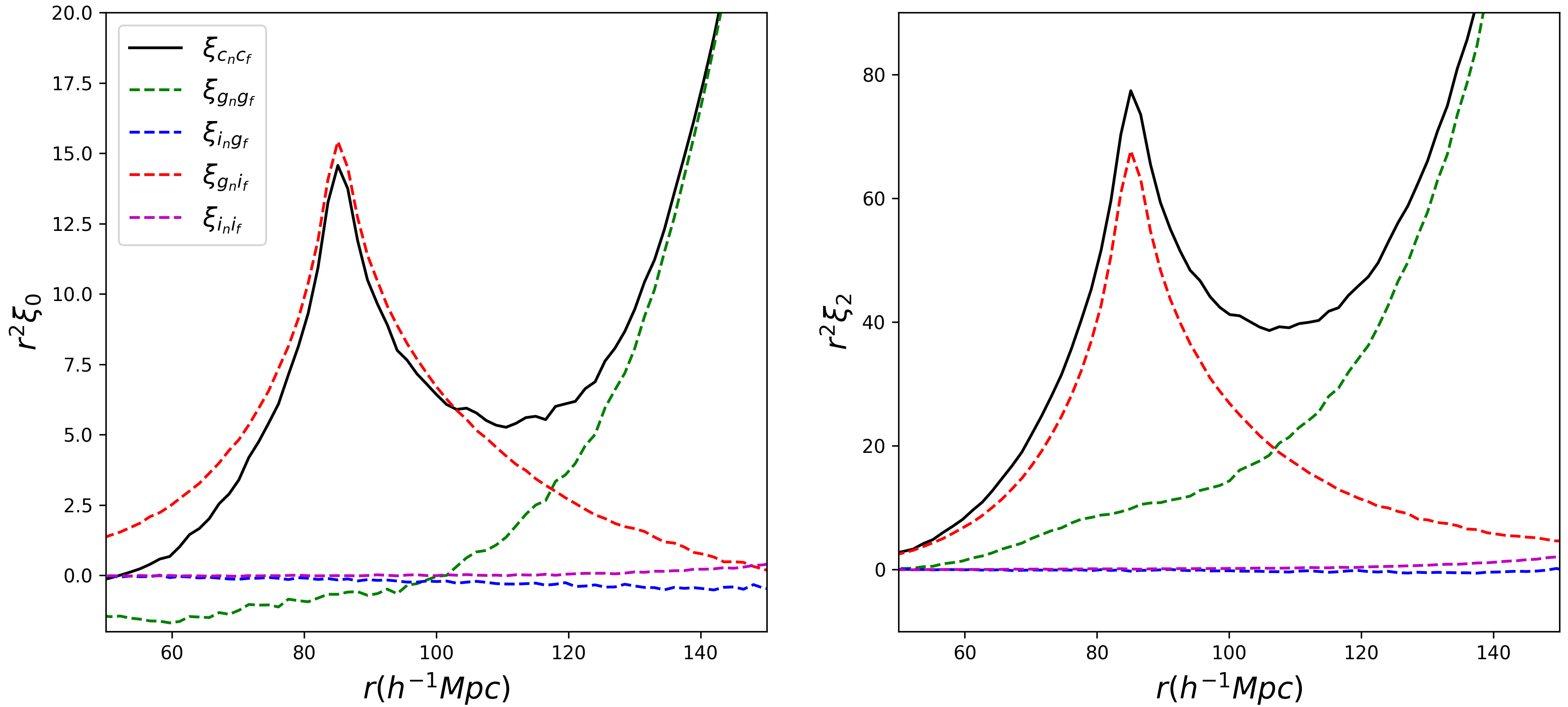}
\caption{Contributions of each term of \ref{eq:ud_corr} to the total near/far-shifted cross correlation $\xi_{c_nc_f}$. The monopole is shown on the left and the quadrupole on the right. \label{fig:components}}
\end{figure}

The key insight that led to this approach is that the near-shifted targets are still separated by the same distance from the far-shifted interlopers as the original interloper-target pairs. Hence the cross term $\xi_{g_{n}i_f}$ in Equation~\ref{eq:ud_corr} is equal to the cross term $\xi_{gi}$ in Equation~\ref{eq:cc_corr}. Note that the direction of the displacement between galaxy targets and interlopers is reversed in $\xi_{g_{n}i_f}$ compared to $\xi_{gi}$ (see Figure~\ref{fig:schematic}). This does not cause any issues, as we are only interested in fitting for even multipoles of the correlation function. Furthermore, this term dominates the near-far cross correlation. Unfortunately, the other terms in Equation~\ref{eq:ud_corr} are not negligible and need to be included in the model. The contributions of each component to Equation~\ref{eq:ud_corr} are shown as dashed colored lines in Figure~\ref{fig:components}, while the full near-far cross correlation is displayed in black.

As anticipated, $\xi_{g_{n}i_f}$ dominates the expression. As the scales and shape of this term match those in the contaminated auto-correlation, and we are interested in fitting BAO on scales $\sim$ 50 - 150 \hinv, these are also the scales for which we need to model the other components. The next most important component is $\xi_{g_ng_f}$, followed by $\xi_{i_{n}i_{f}}$. In these terms, pairs of galaxy targets or interlopers that are at small separation $r_\parallel$ along the LOS in the original contaminated catalogue are mapped to a separation $r_\parallel+2\Delta\sim r_\parallel+200$ $h^{-1} {\rm Mpc}$ in the near-far shifted cross-correlation. Thus most of the small-scale clustering is mapped outside the fitting range. On the other hand, galaxy targets or interlopers that are at a large separation ($\sim2\Delta$) along the LOS in the original contaminated catalogue, become pairs at small to intermediate separation in the near-far shifted cross-correlation. When considering near-far pairs perpendicular to the LOS, those that are at $\sim$ 50 - 150 $h^{-1} {\rm Mpc}$ correspond to objects that are at distance larger than $\sim2\Delta$ in the original contaminated catalogue. Because of this, $\xi_{g_ng_f}$ and $\xi_{i_{n}i_{f}}$ are sourced by large-scale clustering, they have a small amplitude --- $\xi_{i_{n}i_{f}}$ is additionally suppressed by the factor $f_i^2$ --- and they can be modelled using linear theory.  Similarly, the final term $\xi_{i_{n}g_{f}}$ is also the result of large-scale pairs in the contaminated catalogue and has a very small amplitude, making it important only when we work in redshift space, due to the Kaiser effect.

Now returning to Equation~\ref{eq:ud_corr}, we can rearrange to isolate the target-interloper cross term.

\begin{equation}
    f_i(1-f_i) \xi_{g_ni_f}(\Vec{r}) =
    \xi(\Vec{r}, f_{i})_{c_nc_f} 
    - (1-f_i)^2\xi_{g_ng_f}
    - f_i^2\xi_{i_ni_f}
    - f_i(1-f_i)\xi_{i_ng_f}\,.
    \label{eq:gi_cross}
\end{equation}
As $\xi_{gi}$ = $\xi_{g_ni_f}$, we can substitute into Equation~\ref{eq:cc_corr} and form a model for the contaminated correlation function:
\begin{equation}
    \xi_{cc} = (1-f_i)^2\xi_{gg}
    + f_i^2\xi_{ii} + 2\xi_{c_nc_f} 
    - 2(1-f_i)^2\xi_{g_ng_f}
    - 2f_i^2\xi_{i_ni_f}
    - 2f_i(1-f_i)\xi_{i_ng_f}\,.
    \label{eq:cc_with_map}
\end{equation}

We now have a model for the contaminated correlation function using only the measured near/far-shifted cross correlation $\xi_{c_nc_f}$ and the usual large-scale auto-correlation template to describe $\xi_{gg}$, $\xi_{ii}$, and terms that may be modelled with the same auto-correlation.

\subsection{Including interlopers in the BAO model}
\label{sec:ap_rsd}

Before a correlation function or power spectrum may be measured, a fiducial cosmology must be assumed to translate redshifts into physical separations. The BAO technique is based on the peak being shifted if there is a difference between the true cosmology and the assumed fiducial cosmology. This distortion is characterised using the Alcock-Paczynski (AP) parameters $\alpha$ and $\epsilon$ \cite{ap},
\begin{equation}
    \alpha = \left[\frac{D_{\text{A, true}}^{2}(z)H_{\text{fid}}(z)}{D_{\text{A, fid}}^{2}(z)H_{\text{true}}(z)}\right]^{1/3}\frac{r_{\text{s, fid}}}{r_{\text{s, true}}}, \quad
    1+\epsilon = \left[\frac{D_{\text{A, fid}}^{2}(z)H_{\text{fid}}(z)}{D_{\text{A, true}}^{2}(z)H_{\text{true}}(z)}\right]^{1/3}\,,
\end{equation}
where $D_A$ is the angular diameter distance, and $H$ the Hubble parameter. $r_s$ is the comoving sound horizon at the baryon drag epoch. Using these parameters, we may relate the true parallel and perpendicular separations,
\begin{equation}
    r_{\parallel, \text{true}} = \alpha(1+\epsilon)^2 r_{\parallel, \text{fid}}, \quad
    r_{\perp, \text{true}} = \alpha(1+\epsilon)^{-1} r_{\perp, \text{fid}}.
\end{equation}
The fiducial Cartesian coordinates and true polar coordinates can be related as follows,
\begin{equation}
    r_{\text{true}} = \alpha \sqrt{(1+\epsilon)^4 r_{\parallel}^{2} + (1+\epsilon)^{-2} r_{\perp}^{2}}, \quad
    \mu_{\text{true}} = \frac{(1+\epsilon)^2 r_{\parallel}}{\sqrt{(1+\epsilon)^4 r_{\parallel}^{2}  + (1+\epsilon)^{-2} r_{\perp}^{2} }}.
\end{equation}

In order to extract the BAO signal and marginalise over other information in the correlation function, it is standard to fit a model such as 
\begin{equation}
    \xi_{\text{model, $\ell$}}(a_1,a_2,a_3,B,\alpha, \epsilon, r) = \frac{a_{1}}{r^{2}}+\frac{a_{2}}{r} + a_{3}+ B\times \xi_{\text{galaxy, $\ell$}}(\alpha, \epsilon, r)\,,
\end{equation}
where $\xi_{\text{galaxy}}$ is the large-scale auto-correlation template that we will describe in Section~\ref{sec:CAMB_to_xi}, the multiplicative factor $B$ is added to adjust the amplitude, and a polynomial term $A(r)$ is added to account for scale dependant bias and match the overall broadband shape \cite{Ross-BOSS}. $\alpha$ and $\epsilon$ are the Alcock-Paczynski parameters and the $a_i, B$ are nuisance parameters. These nuisance parameters are chosen such that they may modulate the template to match the data, but without the flexibility to fit the BAO peak itself. With this form for the auto-correlation multipoles, we may combine with Equation~\ref{eq:cc_with_map} to construct models for the multipole moments of the contaminated correlation function:
\begin{align}
    \xi_{\text{cc}, \ell}(r)
    &= \frac{a_{1, \ell}}{r^{2}}+\frac{a_{2, \ell}}{r} + a_{3, \ell}+(B_1(1-f_i)^{2} + B_2 f_i^{2} ) \times \xi_{\text{galaxy}, \ell}(\alpha, \epsilon, r) \nonumber \\[0.7em] 
    &+ 2\xi_{c_nc_f, \ell}(r) \label{eq:monomodelbias} \\[0.7em]
    &- 2(B_1(1-f_i)^{2} + B_2 f_i^{2} ) \times  \left(\mathcal{M}_{2\Delta}\left[\xi_{\text{galaxy}}\right]\right)_l(\alpha, \epsilon, r) \nonumber \\[0.7em] 
    &- 2\sqrt{B_2} f_i \sqrt{B_1}(1 - f_i) \times \left(\mathcal{M}_{3\Delta}\left[\xi_{\text{galaxy}}\right]\right)_l(\alpha, \epsilon, r) \nonumber\,,
\end{align}

Here, we have used a mapping shift operator, $\mathcal{M}_X$ introduced in \cite{foroozan2022}, acting on the full two dimensional correlation function $\xi_{\text{galaxy}}$. This operator acts on the argument of a correlation function by mapping a separation $(r, \mu)$ into $(r_{\text{shift}},\mu_{\text{shift}})$ to describe the distortion due to a increase in LOS separation $X$ between the two objects in each pair count.\footnote{Initially pairs have separation $\Vec{r} = \Vec{x_2} - \Vec{x_1}$. After increasing the separation by $\Vec{X}$ (with direction along the LOS), the new separation vector is $\Vec{r} \ ' = \Vec{x_2}' - \Vec{x_1}'$, where $\Vec{x_2}' = \Vec{x_2} + \Vec{X}$ and $\Vec{x_1}' = \Vec{x_1}$. Only one of the galaxy positions are changed.} For any pair with initial separation $(r, \mu)$, the shifted coordinates are given by a simple trigonometric mapping for a LOS shift $X$
\begin{equation}
    r_{\text{shift}}(r, \mu) = \sqrt{r^{2} + X^{2} - 2Xr\mu}, \quad
    \mu_{\text{shift}}(r, \mu) = \frac{r\mu - X}{r_{\text{shift}}(r, \mu)}\,. 
    \label{eq:mapping}
\end{equation}

For example, we have written our unweighted estimate of $\xi_{g_ng_f}$ as $\xi_{g_ng_f} \approx \mathcal{M}_{2\Delta}[\xi_{gg}]$. This takes the auto-correlation of galaxies, at their true positions, and generates the cross-correlation of two catalogues of galaxies separated by 2$\Delta$. In $\xi_{i_ng_f}$, the galaxy target/interloper populations are shifted 2$\Delta$ from their observed positions (as with all terms in the near/far shifted correlation), but the interlopers had been already shifted by 1$\Delta$ from their true positions. Thus, we write the unweighted estimate of $\xi_{i_ng_f}$ as $\xi_{i_ng_f} \approx \mathcal{M}_{3\Delta}[\xi_{gg}]$. 

For the third and fourth lines of Equation~\ref{eq:monomodelbias}, note that the polynomial term described above is not included in the estimate of the auto-correlation. As described in Section~\ref{sec:selfcalib}, these terms are derived from the large-scale correlation, and can be modelled using linear theory. Thus, the polynomial term is not required here.

We have written these equations in a way that allows us to include differing bias schemes for interlopers and targets. We consider that a reasonable estimate is to separate the amplitude parameter $B$ into $B_1$ and $B_2$ to account for two different biases, weighted by the interloper fraction for targets and interlopers. This also allows us to estimate $\xi_{i_{n}g_{f}}$ as we can use the mapping for 3$\Delta$, while adding in factors of $\sqrt{B_1}$ and $\sqrt{B_2}$ (since $B_1$, $B_2$ are the squares of the usual biases) to correctly scale the term along with the others. However, unless otherwise stated, we combine both $B_1$, $B_2$, when fitting mocks with only a single bias.

When fitting both multipoles simultaneously, this results in a model with 11 parameters. When not working in redshift-space, we have found that we can omit the last term in the models containing $\mathcal{M}_{3\Delta}$.

\subsection{Obtaining the auto-correlation template}
\label{sec:CAMB_to_xi}

Now we describe the methods to obtain the template $\xi_{\text{galaxy}}$. We first obtain the linear matter power spectrum $P(k)$ from CAMB using the fiducial cosmology of the Quijote simulations, and produce a real-space matter power spectrum model that takes into account how non-linear effects modify the BAO peak,
\begin{equation}
    P(k,\mu) = [P_{\text{lin}} - P_{\text{smooth}}]e^{-\frac{1}{2}k^2\left(\mu^{2}\Sigma_{\parallel}^{2} + (1-\mu^2)\Sigma_{\perp}^2\right)} + P_{\text{smooth}}\,,
    \label{eq:Pmatter}
\end{equation}
where $P_{\text{smooth}}$ is the linear matter power spectrum smoothed as in \cite{vlah2016} to remove the BAO wiggles. The Gaussian in the first term damps the BAO wiggles ($P_{\text{lin}} - P_{\text{smooth}}$) to account for late-time non-linear evolution. We use $\Sigma_{\perp}$ = 4.8 $h^{-1} {\rm Mpc}$ and $\Sigma_{\parallel}$ = 7.3 $h^{-1} {\rm Mpc}$ evaluated at z = 1 in the fiducial cosmology \cite{cohn2016}. The effect of changing these parameters on the correlation function template is shown in \cite{xu2013baofitting}.

We model the redshift space galaxy power spectrum using a large-scale linear galaxy bias, with the inclusion of the Kaiser effect \cite{kaiser1987} and the Fingers-of-God (FoG) \cite{jackson1972} features, resulting in:
\begin{equation}
    P_{s} = b^{2}(1+\beta\mu^{2})^2 F(k,\mu)P(k,\mu).
    \label{eq:Pk_rsd}
\end{equation}
We use $F(k,\mu) = 1${\color{blue}:} As we are using halos in place of galaxies, we can consider the FoG effect to be negligible. $\beta$ is given by $f/b$, where $f$ is the linear growth rate approximated to $\Omega_m(z)^{0.55}$, and $b$ is the linear galaxy bias that we set to $b=2.8$ being the halo bias of the used simulations. We then take the Fourier transform of this redshift space power spectrum to obtain the two-point correlation functions $\xi_{\rm galaxy}$ and its multipole moments.

We have also considered using analytic templates from HALOFIT \cite{smith2003,takahashi2012} as the real-space matter power spectrum  when modeling $\xi_{\rm galaxy}$ in the additional terms describing the near-far shifted correlation functions in Equation~\ref{eq:monomodelbias} (last two rows) to see if these impact our fits. We find little difference, again a consequence of only needing to model these terms on large scales. 

\section{Testing the method using simulations}
\label{sec:results}

In this section, we present the results of fitting our model to the contaminated catalogues in the simulations. We fit the mean of both the monopole and quadrupole moments of 1000 correlation functions in redshift space via a Markov-chain Monte Carlo (MCMC) analysis using the emcee python package \cite{emcee} and minimizing the following posterior distribution \cite{percival2021}:

\begin{equation}
    f(\xi^{\text{model}}|\xi^{\text{data}}, C) \propto \left[ 1+\frac{\chi^2}{n_s -1}\right]^{-\frac{m}{2}}, 
\end{equation}
where $n_s$ is the number of simulations, $m$ is given by Equation 54 in \cite{percival2021}, and $\chi^2 =(\xi^{\text{model}}-\xi^{\text{data}})^TC^{-1}(\xi^{\text{model}}-\xi^{\text{data}})$ with $C^{-1}$ being the inverse of the covariance matrix

\begin{equation}
    C_{ij}[\xi(r_{i})\xi(r_{j})] = \frac{1}{n_s(n_s - 1)} \sum^{n_s}_{n=1} [\xi_{n}(r_{i}) - \bar{\xi}(r_{i})][\xi_{n}(r_{j}) - \bar{\xi}(r_{j})]
\end{equation}
evaluated for all points in $\xi=\xi^{\text{data}}=\xi_{cc} - 2\xi_{c_nc_f}$. The BAO fit is performed using the monopole and quadrupole within the range $r\in [50,120]~h^{-1}$Mpc.

In order to quantify the performance of our BAO fitting model in Equation~\ref{eq:monomodelbias}, and how this depends on the amount of interlopers in the contaminated catalogue and on the particular value for the interloper displacement (thus the redshift of the survey), we consider 12 different scenarios given by the combination of 3 different values for the interloper displacement $\Delta$ = (85, 90, 97 \hinv) and 4 values for the interloper fraction $f_i$ = (0.02, 0.05, 0.10, 0.15). 

\subsection{Results}
\label{sec:final_results}

\begin{figure}[htbp]
\centering
\includegraphics[width=0.9\textwidth]{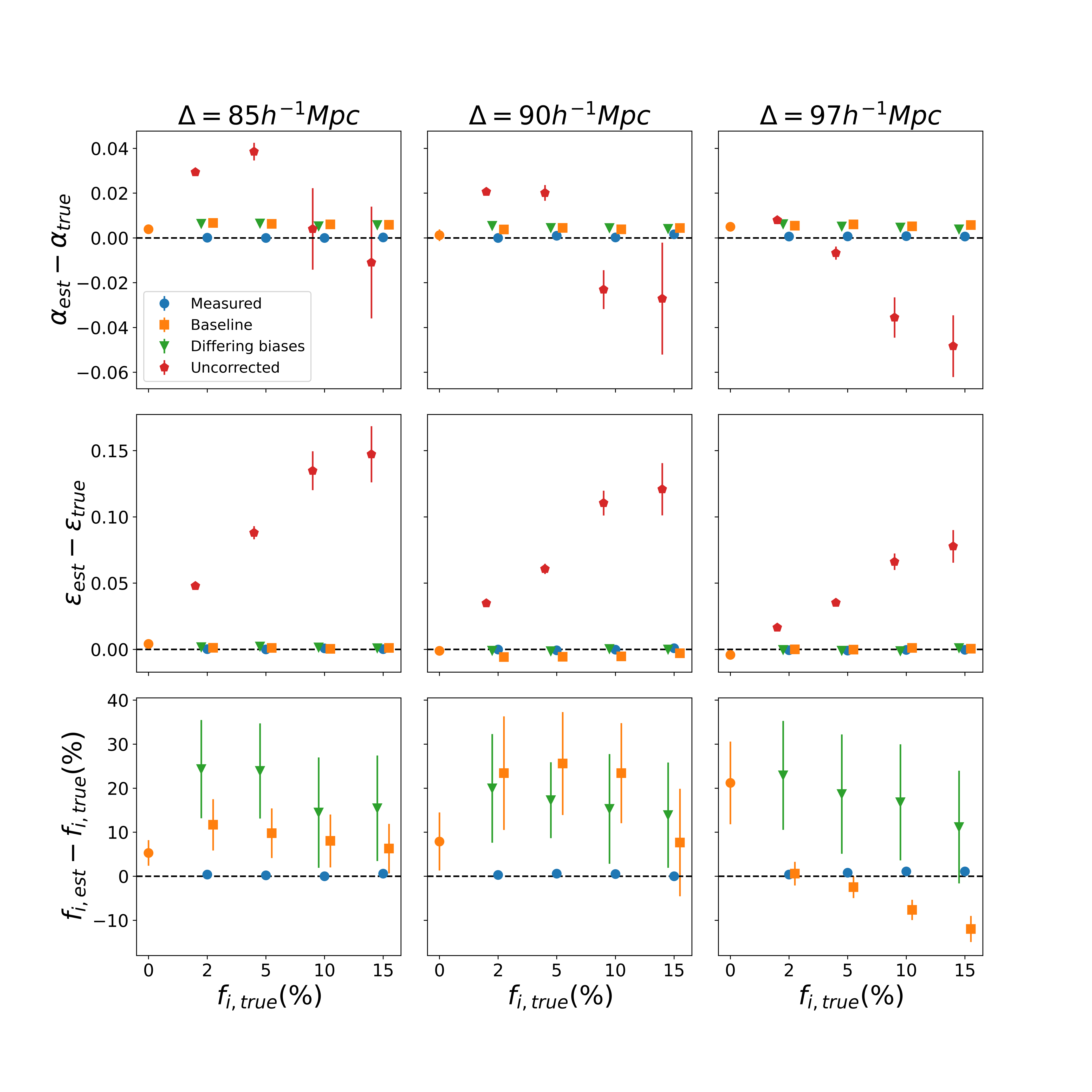}
\caption{ Systematic errors for $\alpha$ (top row), $\epsilon$ (middle row), and $f_i$ (bottom row), plotted against the true fraction of interlopers. The case with no correction for the presence of interlopers is shown with the red pentagons (``Uncorrected"). Fits using the measured galaxy auto-correlation are shown with the blue circles (``Measured''). Fits using an estimated galaxy auto-correlation derived from CAMB are shown with the orange squares (``Baseline''). Fits using mocks where interlopers are more strongly biased are shown in the green triangles ("Differing Biases").}
\label{fig:results}
\end{figure}

First, we performed the BAO fit using a model that does not account for the presence of interlopers (using the first line only in Equation~\ref{eq:monomodelbias}) to quantify how interlopers bias the estimation for the dilation parameters $\alpha$ and $\epsilon$. Our findings are consistent with those in~\cite{massara2020,foroozan2022}: the systematic errors on both dilation parameters increase with the interloper fraction of the contaminated catalogue and they reach maximum values up to  0.048 and 0.15 for $\alpha$ and $\epsilon$, respectively, when $f_i=0.15$. Results for the 12 combinations of interloper fractions and displacements are shown in red in Figure~\ref{fig:results}. In general, bias in the estimates for the AP parameters increases with interloper fraction, and the bias increases faster if the interloper displacement is further from the BAO peak. For example, consider $\Delta = 85$ \hinv, the displacement we consider that is furthest from the BAO peak. The systematic error on $\epsilon$ ranges from about 0.05 to 0.15. However, with $\Delta = 97$ \hinv, much closer to the BAO peak at $\sim$ 100 \hinv, the bias on $\epsilon$ is both smaller at $f_i = 0.02$ and only reaches $\sim 0.07$, spanning only half the range as the case with $\Delta = 85$ \hinv. The average systematic error on all three parameters ($\alpha, \epsilon, f_i$) decreases with increasing interloper displacement $\Delta$, as the interloper peak overlaps more with the BAO peak. For $\Delta > r_{BAO}$, the trend of increasing parameter bias with increasing interloper displacement does not continue, as investigated in \cite{foroozan2022}. As the interloper peak is shifted further away from the BAO peak, it moves outside the fitting range and its effects on the AP parameters are significantly reduced. The peak systematic error in both parameters is obtained when the absolute difference between the positions of the two peaks is $\sim$ 30 \hinv \ \cite{foroozan2022}.

The fitting model requires describing the uncontaminated galaxy correlation function $\xi_{\rm galaxy}$, which involves the use of a linear prediction for the matter power spectrum coupled to a model for the BAO damping to account for the nonlinear evolution around the BAO scale only, and a linear bias scheme to describe the difference between the matter and galaxy clustering. These are approximations: additional parameters, such as amplitude parameters and polynomial terms, are introduced to account for residual nonlinearities. In order to distinguish between any bias caused by these approximations and the novel way in which we model the galaxy target-interloper cross-correlation via Equation~\ref{eq:gi_cross}, we consider a model where $\xi_{\rm galaxy}$ is directly measured in simulations and inserted in all the terms in Equation~\ref{eq:monomodelbias}. Results are shown as blue points in Figure~\ref{fig:results} and in Table \ref{tab:simresults}. The model is accurate and gives unbiased estimates of all three parameters of interest. The majority of systematic errors are within the statistical errors, with the systematic error of all three parameters remaining consistent regardless of interloper fraction or displacement.

We proceed by implementing the full model where $\xi_{\rm galaxy}$ is computed from the CAMB power spectrum and following the procedure explained in Section~\ref{sec:CAMB_to_xi}. These fits are the main results of this paper, and are shown as orange squares in Figure~\ref{fig:results} and they are reported in Table~\ref{tab:rsdresults}. In this case, both the statistical and systematic errors increase compared to using the measured galaxy auto-correlation. In particular, the systematic error is about $0.4-0.6\%$ for $\alpha$, but it is consistent with the systematic error attributed to the standard pre-reconstruction BAO fit pipeline without the presence of interlopers \cite{alam2017}. Estimates for epsilon remain relatively unchanged and generally are still within statistical errors. As expected, we see significantly poorer results for $f_i$ in particular. The case using $\Delta$ = 90 $h^{-1} {\rm Mpc}$ is notably different from the other two, with more biased estimates of $f_i$ and $\epsilon$, and less biased estimates of $\alpha$. With this displacement case, $f_i$ is significantly overestimated regardless of interloper fraction. The opposite is true for $\Delta$ = 97 \hinv, with the interloper fraction underestimated for most interloper fractions. See Table \ref{tab:rsdresults}. We also consider the applying the full model (all lines of Equation~\ref{eq:monomodelbias}) to catalogues with no interlopers. Without any contaminants, there is no signal to be fitted, so the model is free to fit noise, thus $f_i$ is poorly constrained and any degeneracies with the BAO model are become more important. Howevever, both systematic and statistical errors on the AP parameters are unaffected compared to the baseline case and are still consistent with the standard pre-reconstruction BAO fit pipeline without the presence of interlopers \cite{alam2017}. Thus, the application of our model to data without interlopers will not bias other parameters. These points are shown in orange at $f_{i, {\rm true}} = 0$ in Figure~\ref{fig:results} and Figure~\ref{fig:resultsprior}. 

Finally, we consider the catalogues where interlopers have greater bias than galaxy targets. The results are reported in Table \ref{tab:biasresults} and are shown in Figure~\ref{fig:results} as green triangles. In this case, it is necessary to use the two amplitude parameters $B_1$ and $B_2$ to scale for the relative bias between galaxy targets and interlopers. We see increased statistical error on estimates of $f_{i}$, and now this parameter is consistently overestimated by 1-2$\sigma$. This is in better agreement with the true values, although only to the inflated uncertainty. The estimates for $\epsilon$ remain robust regardless of the testing case. 

\section{Discussion}
\label{sec:discussion}
\subsection{Galaxy bias}
\begin{figure}[htbp]
\centering
\includegraphics[width=0.5\textwidth]{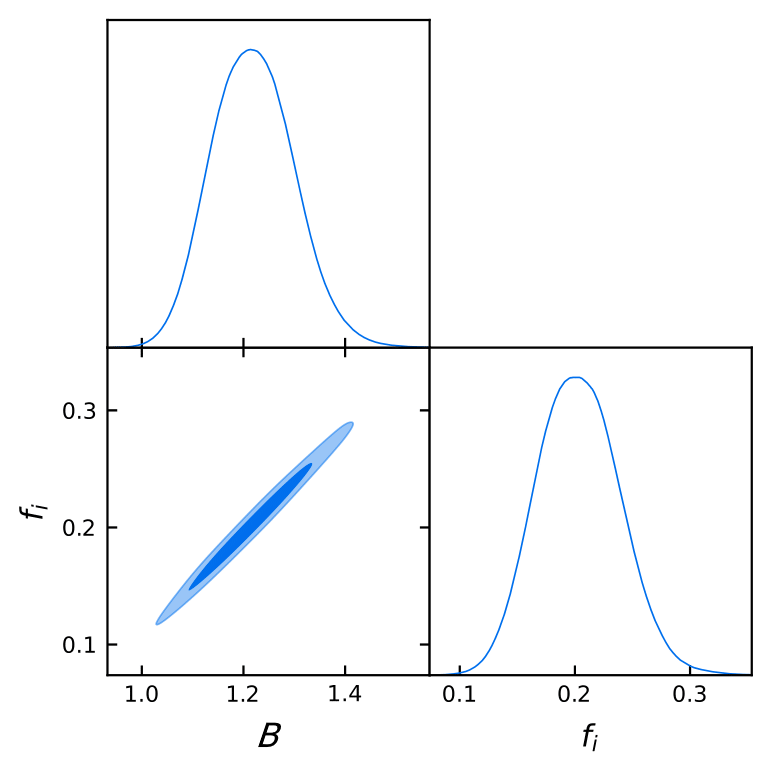}
\caption{The posterior distributions of $B$ and $f_{i}$ from the case using $\Delta$ = 85 $h^{-1} {\rm Mpc}$ and $f_{i, true} = 0.1$. \label{fig:posteriorbf}}
\end{figure}

\begin{figure}[htbp]
\centering
\includegraphics[width=0.9\textwidth]{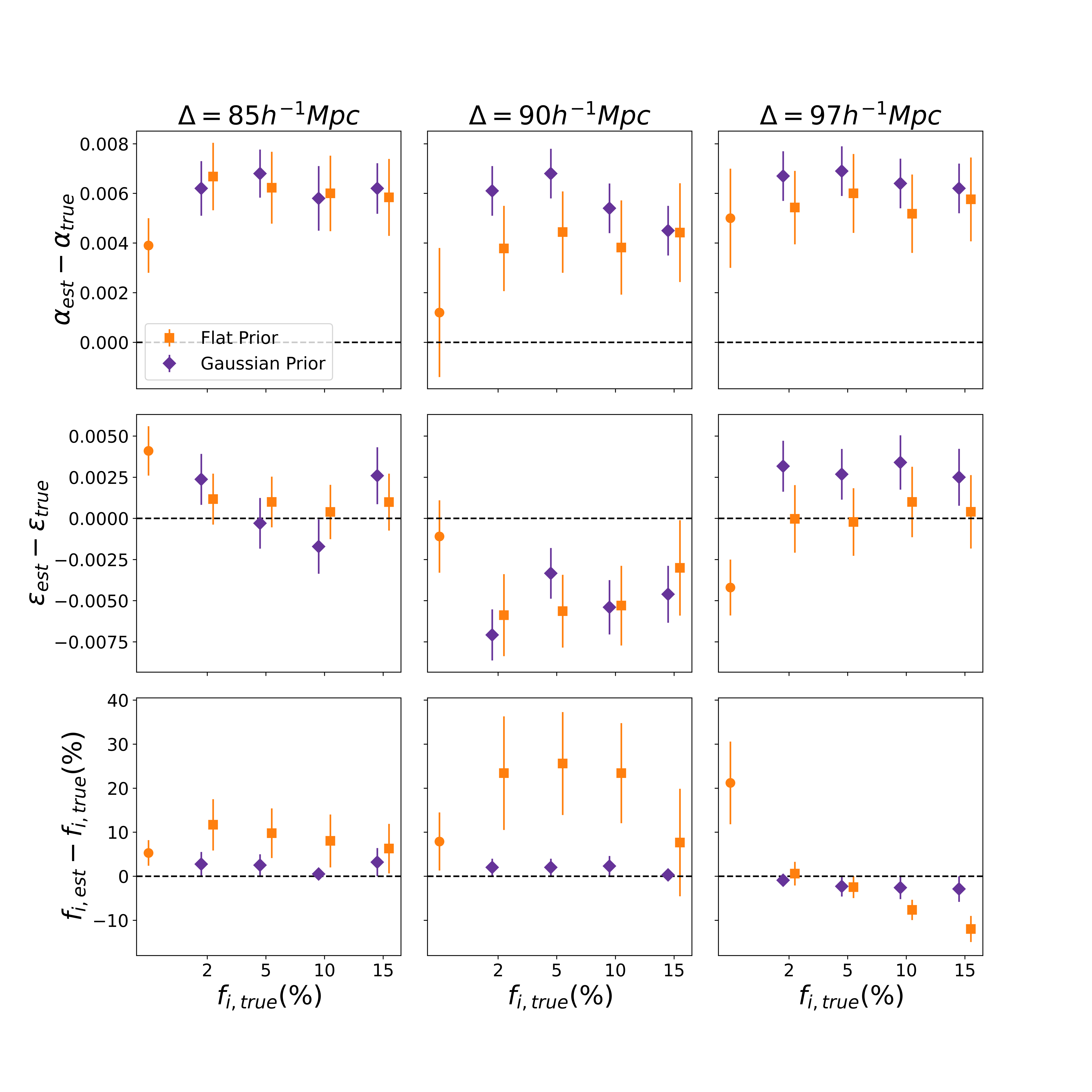}
\caption{Improved parameter estimates applying gaussian prior, shown in purple diamonds, on $B$ for single bias baseline case. Systematic errors for $\alpha$ (top row), $\epsilon$ (middle row), and $f_i$ (bottom row), plotted against the true fraction of interlopers. Vertical scales are the same as Figure~\ref{fig:results}. The improvement in the estimates of $f_i$ is significant. \label{fig:resultsprior}}
\end{figure}

The primary difference between fits using the measured galaxy auto-correlation and using a CAMB template are the estimates of $f_i$. While the estimates using the measured correlation function give unbiased estimates for $f_i$, when CAMB is used, the $B$ parameter is required to adjust the amplitude, and in the case with differing galaxy biases, we require two $B_{i}$. These parameters are highly degenerate with $f_i$ (see Figure~\ref{fig:posteriorbf}), as they are all amplitude-like parameters. It is possible to improve the estimation of $f_i$ when using CAMB by changing the prior on $B$ from a flat to a Gaussian prior centred on $B$ = 1 and with standard deviation = 0.02 to allow for some flexibility. We use a centre of $B$ = 1, as this parameter is not an absolute bias compared to the matter power spectrum. Indeed, when producing the correlation function templates in Section~\ref{sec:CAMB_to_xi}, an estimate of the halo bias $b$ in the Quijote simulations is used. Thus, $B$ serves to correct this estimate, and for a reasonable estimate, we would expect $B$ $\approx$ 1. We show the results using this Gaussian prior in Figure~\ref{fig:resultsprior} in orange. This prior gives estimates that are consistent, within uncertainties, with the AP parameters as compared to the flat prior case (displayed in orange in the same figure).

In the case where target and interloper galaxies exhibit differing biases, it is not as simple to set a similar prior on both $B_1$ and $B_2$. If we consider the Gaussian prior for $B_1$ and $B_2$, set around 1, it needs to be wide enough to capture a sufficient range of possible different biases. This makes no difference compared to a flat prior. Conversely, if we consider offsetting the priors for $B_1$ and $B_2$, this assumes we know which population, targets or interlopers, are more biased, which is an assumption we do not make in this paper. We have tested the effect of knowing the relative bias of targets and interlopers. We performed the fit using the measured galaxy autocorrelation template  with the relative biases, $B_1, B_2$, fixed to their values measured from the mocks. This gives values for $f_i$ well within 1$\sigma$ of the truth. While it is indeed possible to compare the biases of an [OIII] and H$\beta$ objects, interlopers are a subsample of a full H$\beta$ sample - those that do not have [OIII] emission allowing misclassification. Thus, they do not necessarily share the same bias, and accurately estimating the difference in bias is as not straightforward.  The use of future deep fields can help constrain the bias of the interloper subsample and will allow for reasonable priors on $B_1$ and $B_2$.

There is also the possibility of placing a prior on the fraction of interlopers. This, like a prior on $B$ is equivalent to knowing the halo or galaxy bias to some confidence, is equivalent of some knowledge of the expected interloper fraction. Observational estimates are indeed possible using other surveys, for example via [OIII]/H$\beta$ equivalent width ratios measured from the MOSDEF Survey \cite{reddy2018}, giving an expected interloper fraction around 5$\%$ for Roman~\cite{massara2020}. Upcoming deep fields may also be used to place constraints on the expected $f_i$. It is then trivial to place a Gaussian prior on the interloper fraction within our method.

\subsection{Limitations of the Simulation Box}

While we work in a simulation box with periodic boundary conditions, the total number of pairs in $\xi_{c_nc_f}$ is unchanged compared to $\xi_{cc}$. All pairs with the same initial separation will have the same new separation after the shift, and the information contained in both correlation functions is the same. However, in the near-far cross-correlation we have changed the relative scales at which the different components overlap, which allows us to more cleanly separate and model them to extract the cross-correlation term.

In a survey volume, we cannot apply a periodic boundary condition as in the simulation box. Instead, we would need to additionally shift the randoms used to determine the expected density and the correlation function. Thus, the pair counts at all scales should be compared to their expected counts in a random field, without missing the correct normalizations for the large scale correlation we introduce via the near/far shift.

There will be a number of other complications when moving to a realistic lightcone survey. First, the assumption of a constant redshift across the simulation box is not true for a survey lightcone. As the interloper displacement is redshift dependent, it varies across the survey and the measured interloper peak appears broadened. Second, our method requires shifting the catalogue by exactly the interloper displacement to obtain the near/far catalogues. Applying a single shift (e.g. given by the average survey redshift) will result in a distorted estimate for the galaxy target-interloper cross term. This effect may be somewhat mitigated by performing the analysis in fine redshift slices. Under a sufficiently fine binning, the assumption of a single redshift can hold for each bin. Moreover, when shifting a lightcone catalogue, the resulting near/far catalogues are subjected to a geometric distortion due to a non-parallel LOS across the survey: The catalogues' volume changes as well as the angular distance between objects and the angular BAO scale. Therefore, the underlying correlation function of the near shifted catalogue is not the same as that of the far shifted catalogue and the cross-correlation between these samples will result in a distorted correlation function. All these effects will need to be carefully quantified and taken into account when deploying our method to surveys.

\subsection{Testing for additional systematics}

We now compare how the statistical and systematic errors change between our baseline model including the near-far derived correction for interlopers, with results of fits using the following data inputs and covariances.

\begin{itemize}

    \item \emph{The Noiseless Near/Far Correlation} ($\xi_{gg}$, $C_{gg}$)

    Here, as opposed to considering $\xi_{cc} - 2\xi_{c_nc_f}$ for each simulation, we now consider subtracting the mean $\overline{\xi}_{c_nc_f}$ from $\xi_{cc}$ for each simulation. For this, the covariance should be that calculated from only $\xi_{cc}$. In practice, we consider a single fit to the mean over all simulations rather than fitting to each, and so this method only differs in the covariance matrix used. Fitting with $C_{cc}$ results in systematic errors for the AP parameters that are nearly identical to our baseline, but with statistical errors that are on average 30\% smaller for interloper fractions of 2\% - 15\%, with a weak dependence on fraction. Thus, including the near/far shifted measurements in the covariance does not bias the results, but it does increase the uncertainty.
    
    \item \emph{Comparison to standard BAO template fit} ($\xi_{gg}$, $C_{gg}$)

    We also compare our results for the AP parameters to a fit on a dataset without interlopers. In this case, the data vector is $\xi_{gg}$, the auto-correlation of galaxies, with associated covariance $C_{gg}$, and the fitting model is given by the first line of Equation~\ref{eq:monomodelbias} with $f_i$ set to zero. The systematic error for $\alpha$ and $\epsilon$ in this case is consistent with the ones obtained using our method applied to contaminated catalogues, while the statistical errors on the AP parameters decreased by approximately 25\%-40\% compared to interloper fractions of 2\%-15\%. Therefore, the systematic error in our method is due to the standard BAO modelling procedure.
    
\end{itemize}

\subsection{Dipole of the near/far cross-correlation}

In a 2D correlation function, a non-zero dipole arises when the function is not invariant under reflection symmetry along the LOS or, equivalently, under a change in the sign of $\mu$, the cosine of the angle between the LOS and the direction connecting each pair of objects. Changing the sign of $\mu$ corresponds to considering the pair AB instead of the pair BA. When measuring a cross-correlation function, one selects two different populations, then chooses which one appears first in the pair counting, and separations and angles between the first and second groups are determined for each pair. In an auto-correlation, this choice is not made. The near/far shifted correlation function is a cross-correlation between the same catalogue shifted in two opposite directions, so that the first object in the pairs always belongs to the near-shifted catalogue, and the second object always belong to the far-shifted catalogue. Thus we expect a dipole in the near/far contaminated correlation function. 

This dipole is due to two effects. First, an artificial dipole appears into the near/far shifted correlation because that is a cross-correlation between the same catalog shifted along the LOS in two different directions. The shift creates an artificial dipole in the contaminated clustering through the terms $\xi_{g_ng_f}$ and $\xi_{i_ni_f}$. The near/far dipole also has features that are due to the presence of interlopers in the contaminated catalogue. 

\begin{figure}[htbp]
\centering
\includegraphics[width=\textwidth]{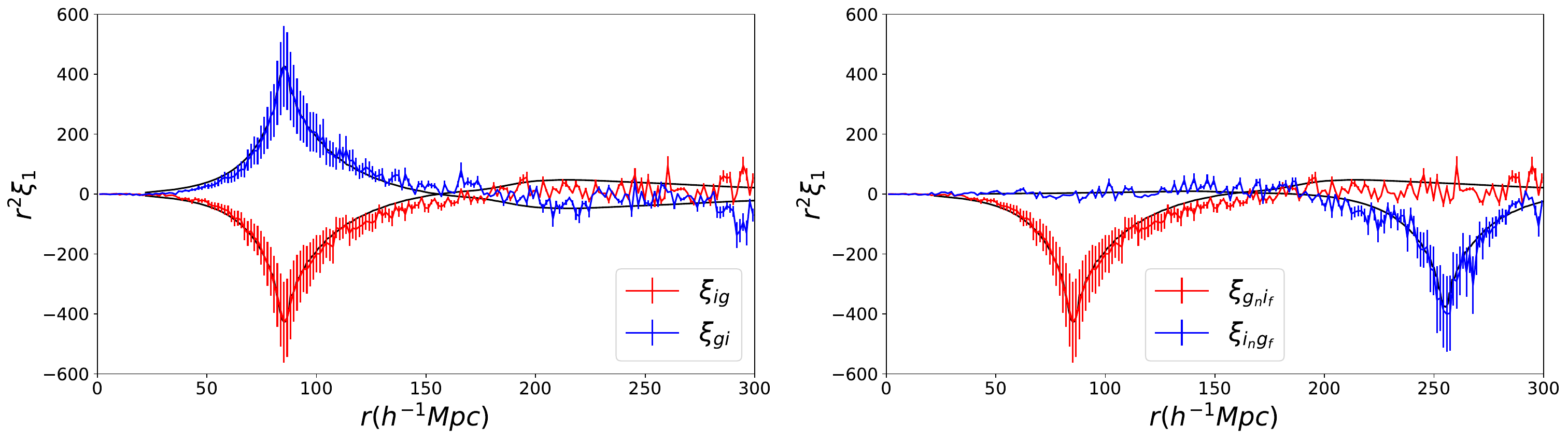}
\caption{The left panel shows measured dipole in the terms $\xi_{ig}$ and $\xi_{gi}$ of the contaminated correlation function (Equation~\ref{eq:cc_corr}). The right panel displays the measured in the terms $\xi_{i_ng_f}$ and $\xi_{g_ni_f}$ in the near-far shifted correlation function (Equation~\ref{eq:ud_corr}). Measurements taken as the mean of 10 mocks. Predictions are shown as the solid black lines in both panels.}
\label{fig:dipole}
\end{figure}

In the contaminated correlation function, the 2D galaxy target-interloper cross correlation can be obtained via $\xi_{gi}\sim\mathcal{M}_\Delta[\xi_{gg}]$, where we applied the transformation $\mathcal{M}$ in Equation~\ref{eq:mapping}. In that transformation, the change $\mu\rightarrow-\mu$ is equivalent to the the change of variable $\Delta\rightarrow-\Delta$. Therefore $\xi_{ig}\sim\mathcal{M}_{-\Delta}[\xi_{gg}]$. While the two cross correlations $\xi_{gi}$ and $\xi_{ig}$ exhibit equal even-multipole components, they present different odd terms. The left panel in Figure~\ref{fig:dipole} shows the dipole of the two cross correlations, as predicted via the $\mathcal{M}_{\pm\Delta}[\xi_{gg}]$ mappings as well as the mean of measurements over 10 mocks. The dipoles have the same amplitude but opposite sign as a function of separation $r$, thus they sum to zero. Since we cannot distinguish between interlopers and target galaxies in observed surveys, these two terms cannot be measured separately and the dipole cannot be detected. 

When considering the near/far shifted correlation function, the dipoles in $\xi_{i_ng_f}$ and $\xi_{g_ni_f}$ appear at different scales, as a result of having shifted catalogues. Indeed, $\xi_{g_ni_f}\sim\mathcal{M}_\Delta[\xi_{gg}]$ and $\xi_{i_ng_f}\sim\mathcal{M}_{3\Delta}[\xi_{gg}]$, making all their multipoles (even and odd) differ as a function of separation $r$. The right panel of Figure~\ref{fig:dipole} shows the dipoles in $\xi_{i_ng_f}$ and $\xi_{g_ni_f}$, which appear at different scales and do not sum to zero as in the contaminated correlation function. Additionally, they now share the same sign, another consequence of shifting the contaminated catalogue to generate the near and far shifted ones. 

\begin{figure}[htbp]
\centering
\includegraphics[width=0.7\textwidth]{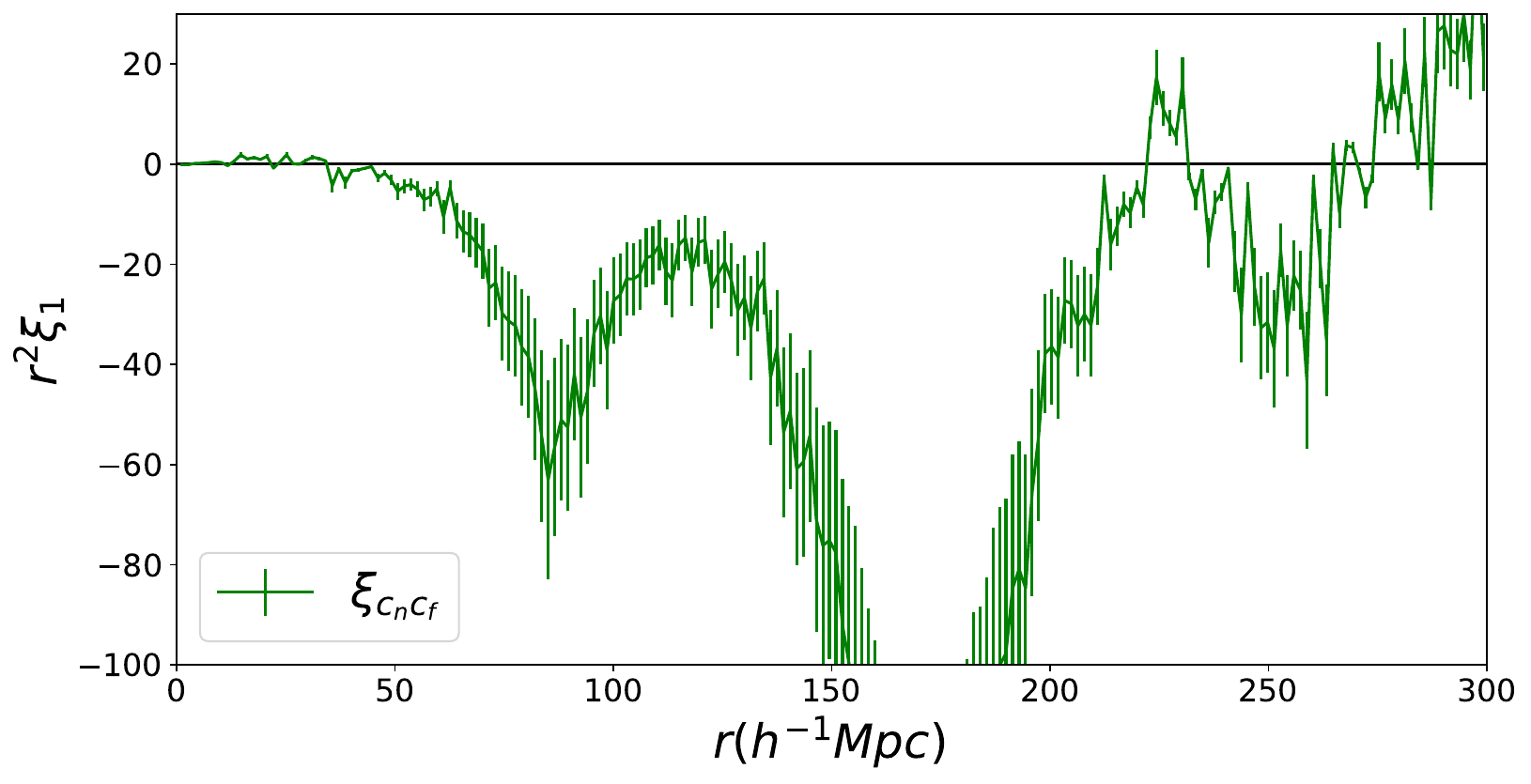}
\caption{The measured dipole of the total $\xi_{c_nc_f}$ is shown in green. Measurements are taken as the mean of 10 mocks. The vertical scale has been limited to a range of [-100, 30], so that the smaller features are more easily seen. The central feature peak is at $\sim$ -1500 and is not shown on this scale.}
\label{fig:dipole_cncf}
\end{figure}

When measuring the near-far shifted correlation function of the full contaminated sample, one will always observe a dominant artificial dipole at exactly the shift scale, $\Delta_{\rm shift}$ as in the monopole. Measurements over 10 mocks of the total near/far shifted correlation dipole are shown in Figure~\ref{fig:dipole_cncf}. In the presence of interlopers, additional offset dipole features will appear. As a result, the dipole can therefore be used as a initial indicator for the presence of this type of interlopers. In fact, the shift need not be equal to some integer multiple of the interloper shift to see this effect. In a situation where the interloper shift is not known, one could perform the near/far shift by any arbitrary value and the additional interloper features would appear at $\Delta_{\rm shift} - \Delta_{\rm interloper}$ and $\Delta_{\rm shift} + \Delta_{\rm interloper}$.

\section{Conclusions}
\label{sec:conclusion}

In this paper, we have discussed how to extract the BAO scale from a galaxy target catalogue that is contaminated with a sample of small-displacement interlopers. This is important for future spectroscopic telescope surveys, like Roman, where it is possible that the population of [OIII] galaxies is contaminated by H$\beta$ interlopers due to single-line redshift measurements. We now cover some important points of our analysis.

The contaminated correlation function has two important terms: The auto-correlation term (which may be split into a galaxy auto-correlation and interloper auto-correlation) and a galaxy target-interloper cross correlation term as a linear combination of two correlation functions. It is this auto-correlation term that we wish to isolate and measure, and thus we sought to model the cross-correlation. 

We have introduced a self-calibration method to estimate this cross term. This involves taking the original contaminated catalogue and creating two new "far-shifted" and "near-shifted" catalogues, with LOS positions shifted towards or away from the observer by one interloper shift. The cross correlation of these catalogues can partially isolate the desired galaxy target-interloper cross term. 

We tested our model in various scenarios exhibiting different values for the interloper fraction and the interloper displacement $\Delta$. In all cases, our model returns unbiased predictions for the AP dilation parameters with systematic errors less than 6.6$\times10^{-3}$ in the worst case, which is consistent with prior pre-reconstruction fitting methods that did not consider catalogues contaminated by interlopers \cite{alam2017}. If interlopers are present but not accounted for in the model, the systematic errors on the dilation parameters are one to two orders of magnitude greater, up to 0.048 and 0.15, for $\alpha$ and $\epsilon$, respectively.

As our analysis will be useful for Roman [OIII] targets, we may predict the statistical error our method would give for a Roman survey. Using a simplistic approach, we may scale the errors we obtain from the effective volume of 1000 mocks (1 $(h^{-1} \text{Gpc})^{3}$ each) to the approximate volume of a survey between z = 1.8 - 2.8, $\sim$ 10 $(h^{-1} \text{Gpc})^{3}$. We are therefore testing the method at a precision a factor of 10 times better than expected from Roman. For the range of interloper fractions and displacements we consider, the systematic errors are therefore expected to be subdominant. The effect of the self-calibration method to ameliorate the effects of interlopers, is that the statistical errors on the AP parameters increase by approximately 30\%, which is only weakly dependent on the interloper fractions for a range 2\% - 15\%. This increase in the errors can be reduced further by providing prior information on the interlopers, for example from deep field observations.

It should also be noted that the change from a simulation box mock to survey data will require additional testing. There are effects due to varying redshift over the survey volume, and the lightcone geometry. These effects should be carefully taken into account before applying the model to real data.

Small-displacement interlopers create a strong signal in the two point correlation function that appears near the BAO peak and thus contaminates measurements of cosmological parameters. With our self-calibration model, it is possible to robustly estimate the AP dilation parameters regardless of differing bias schemes of interlopers and targets. Furthermore, our model can estimate the fraction of interlopers to within a few percent, given the parameters $B$, or $B_1, B_2$, are well known.

\appendix
\section{Tables of values}
\label{tableuncorrectedcov}

The following tables of values were used to generate Figure~\ref{fig:results}. Table~\ref{tab:simresults} contains values for the fit using a measured galaxy auto-correlation function and mocks where galaxy targets and interlopers are equally biased. Table~\ref{tab:rsdresults} contains values for the fit using a galaxy correlation function calculated from CAMB as outlined in Section~\ref{sec:CAMB_to_xi}, and unbiased mocks. Table~\ref{tab:biasresults} contains values for the fit using mocks where interlopers are more strongly biased. 

\begin{table}[htbp]
\centering
\begin{tabular}{c|c|c|c|c|c|c}
\hline
f$_{i}$$^{\text{true}}$($\times$10$^{2}$) & \multicolumn{2}{c}{Error on $\alpha$($\times$10$^{3}$)} & \multicolumn{2}{c}{Error on $\epsilon$($\times$10$^{3}$)} & \multicolumn{2}{c}{Error on f$_{i}$($\times$10$^{2}$)}\\
\hline
\hline
$\Delta$=85&Systematic&Statistical& Systematic & Statistical & Systematic & Statistical\\
\hline
2 & 0.1 & 0.7 & 0.2 & 0.6 & 0.4 & 0.3\\
5 & 0.0 & 1.0 & 0.0 & 0.6 & 0.2 & 0.3\\
10 & 0.0 & 1.0 & 0.7 & 0.8 & 0.0 & 0.3\\
15 & 0.2 & 0.7 & 0.2 & 0.6 & 0.6 & 0.3\\
\hline
\hline
$\Delta$=90&Systematic&Statistical& Systematic & Statistical & Systematic & Statistical\\
\hline
2 & 0.0 & 0.5 & -0.1 & 0.5 & 0.3 & 0.5\\
5 & 1.0 & 1.0 & -0.7 & 0.6 & 0.6 & 0.5\\
10 & 0.2 & 0.7 & -0.2 & 0.7 & 0.5 & 0.5\\
15 & 1.6 & 0.8 & 0.8 & 0.6 & 0.0 & 0.6\\
\hline
\hline
$\Delta$=97&Systematic&Statistical& Systematic & Statistical & Systematic & Statistical\\
\hline
2 & 0.6 & 0.8 & -0.6 & 0.9 & 0.4 & 0.5\\
5 & 0.7 & 0.7 & -0.9 & 1.0 & 0.8 & 0.5\\
10 & 0.8 & 0.9 & -0.4 & 1.0 & 1.1 & 0.6\\
15 & 0.6 & 0.9 & -0.3 & 1.0 & 1.1 & 0.6\\
\hline
\end{tabular}
\caption{Systematic and statistical errors on $\alpha$, $\epsilon$, $f_i$ from multipole fits on the mean of 1000 redshift space mocks and using measured autocorrelation. $\Delta$ given in \hinv. These are the blue circles labelled “Measured” in Figure~\ref{fig:results}.}\label{tab:simresults}
\end{table}

\begin{table}[htbp]
\centering
\begin{tabular}{c|c|c|c|c|c|c}
\hline
f$_{i}$$^{\text{true}}$($\times$10$^{2}$) & \multicolumn{2}{c}{Error on $\alpha$($\times$10$^{3}$)} & \multicolumn{2}{c}{Error on $\epsilon$($\times$10$^{3}$)} & \multicolumn{2}{c}{Error on f$_{i}$($\times$10$^{2}$)}\\
\hline
\hline
$\Delta$=85&Systematic&Statistical& Systematic & Statistical & Systematic & Statistical\\
\hline
0 & 3.9 & 1.1 & 4.1 & 1.5 & 5.3 & 2.9 \\
2 & 6.7 & 1.4 & 1.2 & 1.6 & 11.7 & 5.8 \\
5 & 6.2 & 1.4 & 1.0 & 1.5 & 9.8 & 5.6 \\
10 & 6.0 & 1.5 & 0.4 & 1.6 & 8.0 & 6.0 \\
15 & 5.8 & 1.6 & 1.0 & 1.7 & 6.3 & 5.6 \\
\hline
\hline
$\Delta$=90&Systematic&Statistical& Systematic & Statistical & Systematic & Statistical\\
\hline
0 & 1.2 & 2.6 & -1.1 & 2.2 & 7.9 & 6.6 \\
2 & 3.8 & 1.7 & -5.9 & 2.5 & 23.4 & 12.9 \\
5 & 4.4 & 1.6 & -5.6 & 2.2 & 25.6 & 11.7 \\
10 & 3.8 & 1.9 & -5.3 & 2.4 & 23.4 & 11.4 \\
15 & 4.4 & 2.0 & -3.0 & 2.9 & 7.7 & 12.2 \\
\hline
\hline
$\Delta$=97&Systematic&Statistical& Systematic & Statistical & Systematic & Statistical\\
\hline
0 & 5.0 & 2.0 & -4.4 & 1.7 & 21.2 & 9.4 \\
2 & 5.4 & 1.5 & -0.0 & 2.1 & 0.6 & 2.7 \\
5 & 6.0 & 1.6 & -0.2 & 2.1 & -2.5 & 2.5 \\
10 & 5.2 & 1.6 & 1.0 & 2.1 & -7.6 & 2.3 \\
15 & 5.8 & 1.7 & 0.4 & 2.2 & -12.0 & 3.0 \\
\hline
\end{tabular}
\caption{Systematic and statistical errors on $\alpha$, $\epsilon$, $f_i$ from multipole fits on the mean of 1000 redshift space mocks and using CAMB autocorrelation. $\Delta$ given in \hinv. These are the orange squares labelled ``Baseline" in Figure~\ref{fig:results} and ``Flat Prior" in Figure~\ref{fig:resultsprior}.}\label{tab:rsdresults}
\end{table}

\begin{table}[htbp]
\centering
\begin{tabular}{c|c|c|c|c|c|c}
\hline
f$_{i}$$^{\text{true}}$($\times$10$^{2}$) & \multicolumn{2}{c}{Error on $\alpha$($\times$10$^{3}$)} & \multicolumn{2}{c}{Error on $\epsilon$($\times$10$^{3}$)} & \multicolumn{2}{c}{Error on f$_{i}$($\times$10$^{2}$)}\\
\hline
\hline
$\Delta$=85&Systematic&Statistical& Systematic & Statistical & Systematic & Statistical\\
\hline
2 & 6.3 & 1.5 & 1.7 & 1.7 & 24.3 & 11.2 \\
5 & 6.4 & 1.6 & 2.1 & 1.7 & 23.9 & 10.8 \\
10 & 5.2 & 1.6 & 1.5 & 1.8 & 14.5 & 12.5 \\
15 & 5.8 & 1.7 & 0.9 & 1.8 & 15.5 & 12.0 \\
\hline
\hline
$\Delta$=90&Systematic&Statistical& Systematic & Statistical & Systematic & Statistical\\
\hline
2 & 5.4 & 1.6 & -1.0 & 1.8 & 20.0 & 12.3 \\
5 & 4.5 & 1.8 & -1.4 & 1.9 & 17.3 & 8.6 \\
10 & 4.4 & 1.6 & 0.3 & 2.0 & 15.3 & 12.4 \\
15 & 4.0 & 1.8 & -0.2 & 1.8 & 13.9 & 11.9 \\
\hline
\hline
$\Delta$=97&Systematic&Statistical& Systematic & Statistical & Systematic & Statistical\\
\hline
2 & 6.2 & 1.7 & -0.4 & 2.1 & 22.9 & 12.4 \\
5 & 5.1 & 1.7 & -1.0 & 2.4 & 18.7 & 13.6 \\
10 & 4.7 & 1.8 & -1.4 & 2.5 & 16.8 & 13.2 \\
15 & 3.8 & 1.8 & 0.9 & 2.2 & 11.2 & 12.8 \\
\hline
\end{tabular}
\caption{Systematic and statistical errors on $\alpha$, $\epsilon$, $f_i$ from multipole fits on the mean of 1000 redshift space mocks interlopers more strongly biased. and using CAMB autocorrelation. $\Delta$ given in \hinv. These are the green triangles labelled ``Differing Biases” in Figure~\ref{fig:results}.}\label{tab:biasresults}
\end{table}

\acknowledgments

All authors acknowledge the support of the Canadian Space Agency. WP also acknowledges support from the Natural Sciences and Engineering Research Council of Canada (NSERC), [funding reference number RGPIN-2019-03908].

Research at Perimeter Institute is supported in part by the Government of Canada through the Department of Innovation, Science and Economic Development Canada and by the Province of Ontario through the Ministry of Colleges and Universities.

This research was enabled in part by support provided by Compute Ontario (computeontario.ca) and the Digital Research Alliance of Canada (alliancecan.ca). 

All authors thank Yun Wang for useful discussions. We thank the referee for their helpful suggestions.

\printbibliography

\end{document}